\documentclass[journal=jacsat,manuscript=article]{achemso}
\usepackage{graphicx}
\usepackage{dcolumn}
\usepackage{bm}
\usepackage{amsmath}
\usepackage{amssymb}
\usepackage{ulem}
\usepackage[T1]{fontenc}
\usepackage[section]{placeins}

\usepackage[dvipsnames]{xcolor}
\usepackage{bm,graphicx,hyperref}
\newcommand{\be}{\begin{equation}}
\newcommand{\ee}{\end{equation}}
 
\title{Edge-Driven Phase Transitions in 2D Ice}
\author{Suchit Negi}
\affiliation{Centre for Advanced 2D Materials, National University of Singapore, 117546 Singapore}
\alsoaffiliation{Institute for Functional Intelligent Materials,National University of Singapore, 117544, Singapore}
\author{Alexandra Carvalho}
\affiliation{Centre for Advanced 2D Materials, National University of Singapore, 117546 Singapore}
\alsoaffiliation{Institute for Functional Intelligent Materials,National University of Singapore, 117544, Singapore}
\email{carvalho@nus.edu.sg}
\author{Maxim Trushin}
\affiliation{Centre for Advanced 2D Materials, National University of Singapore, 117546 Singapore}
\alsoaffiliation{Institute for Functional Intelligent Materials,National University of Singapore, 117544, Singapore}
\alsoaffiliation{Department of Materials Science Engineering, National University of Singapore, 117575 Singapore}
\author{A. H. Castro Neto}
\affiliation{Centre for Advanced 2D Materials, National University of Singapore, 117546 Singapore}
\alsoaffiliation{Institute for Functional Intelligent Materials,National University of Singapore, 117544, Singapore}
\alsoaffiliation{Department of Materials Science Engineering, National University of Singapore, 117575 Singapore}
\keywords{two-dimensional, water, 2D, graphene, boron nitride, structure}

\keywords{Boron nitride, water ice, edges, electric field, confinement,  two-dimensional}

\begin{document}

\begin{abstract}
Two-dimensional (2D) water, confined by atomically flat layered materials, may transit into various crystalline phases even at room temperature.
However, to gain full control over the crystalline state, we should 
not only confine water in the out-of-plane direction but also restrict its in-plane motion, forming 2D water clusters or ribbons.
One way to do this is by using an electric field, in particular the intrinsic electric field of an adjacent polar material.
We have found that the crystalline phases of 2D water clusters placed between two hexagonal boron nitride ($h$-BN) nanoribbons are crucially determined by the nanoribbons’ edges,
the resulting polarity of the nanoribbons, and their interlayer distance.
We make use of density functional theory (DFT) with further assistance of molecular dynamics (MD) simulations to establish the comprehensive 
phase diagrams demonstrating transitions between liquid and solid phases and between the states of different crystalline orders.
We also show that the crystalline orders are maintained when water flows between $h$-BN channels under external pressure. 
Our results open a promising pathway towards the control of water structure and its flow by the use of the microscopic electric field of polar materials.

\end{abstract}    

\maketitle

Water has always been a subject of interest in account of its unusual properties and its universal importance across the fields of science and engineering. Its phase diagram contains at least 18 ice phases in addition to the most known $I_h$ ice\cite{salzmann2021structure}, with some of them still being object of research. 
Besides, water can be supercooled and undergo glass transitions to  `glassy' amorphous water states\cite{mishima1998relationship}. 
Moderately confined water, similar to supercooled water, is prevented from crystallizing at low temperatures\cite{ricci2009similarities,cerveny2016confined}, and has long been object of research due to its importance for biological\cite{acosta2015filtering}, pharmaceutical and food industry applications. Most recently, however, water confined at the nanometer scale has been shown instead to display the long-range order characteristic of an `ice' even at room temperature\cite{algara2015square,zangi2003monolayer,gao2018phase,zhu2015compression,jiang2021first,neek2016commensurability,qiu2015water}.

Nanoconfined water is usually studied either in carbon nanotubes \cite{NatureNano2017,kofinger2008macroscopically} realizing one-dimensional (1D) confinement
or in graphene slits squeezing water down to a two-dimensional (2D) monolayer \cite{algara2015square,radha2016molecular,zangi2003monolayer,gao2018phase,raju2018phase,chen2016two}.
The flat geometry of the latter makes it possible to impose an additional in-plane confinement and study 
transition not only between bulk and 2D states but also between 2D and 1D limits by shifting the boundaries of the channel.
The edge effects are then of ultimate importance.

When a polar crystal such as $h$-BN is finite, it can have a permanent dipole moment, depending on its geometry.
As we will describe in the present article, monolayer 2D water can exhibit different crystalline phases if it is confined by {\em finite} boron nitride ribbons.

Most of the 2D ice forms predicted so far have no macroscopic electric dipole,\cite{zangi2003monolayer,gao2018phase,raju2018phase,chen2016two} similar to 3D ices, which are typically anti-ferroelectric, with exception of the polar phases XI and XIX\cite{salzmann2021structure}.
In this article, we show that 2D water confined by finite, polar $h$-BN nanoribbons has at least two polar ice phases that are stable at room temperature. 

We start by examining the energetics of water confined between finite layers using DFT calculations. We then use equilibrium MD simulations to describe the structure of 2D water between finite or infinite $h$-BN or graphene sheets in a wider range of conditions, and offer a qualitative explanation of the edge effect along with an effective electrostatic model.
Finally, we examine the structure of water in stationary flow.

\section{\textcolor{black}{Energetics of Water confined in 1D$\leftrightarrow$2D channels}\label{DFT}}

\color{black}
Previous studies of 2D water have usually assumed a structure that is periodic along both in-plane directions.
However, in reality both the confining layers and the water in between are finite and have edges. We can use this effect to our advantage as a mean to further control the structure of the water.

We will consider a water channel formed by two semi-infinite graphene or boron nitride layers. The channel's in-plane width ($W$), along the $x$ direction is assumed to be infinite (periodic),
and the channel's height (the interlayer distance $D$), along the $y$ direction, is a few \AA.
If the distance between edges (i.e. the channel's length $L$, along the $z$ direction) was comparable to the size of the water molecules, the system would become 1D.
In contrast, if $L$ was infinite, the system would become 2D. We study systems with $10<L<400$~\AA, which are at the 1D$\rightarrow$2D transition.

\normalcolor

\begin{figure}[h!]
\centering
\includegraphics[width=0.5\textwidth]{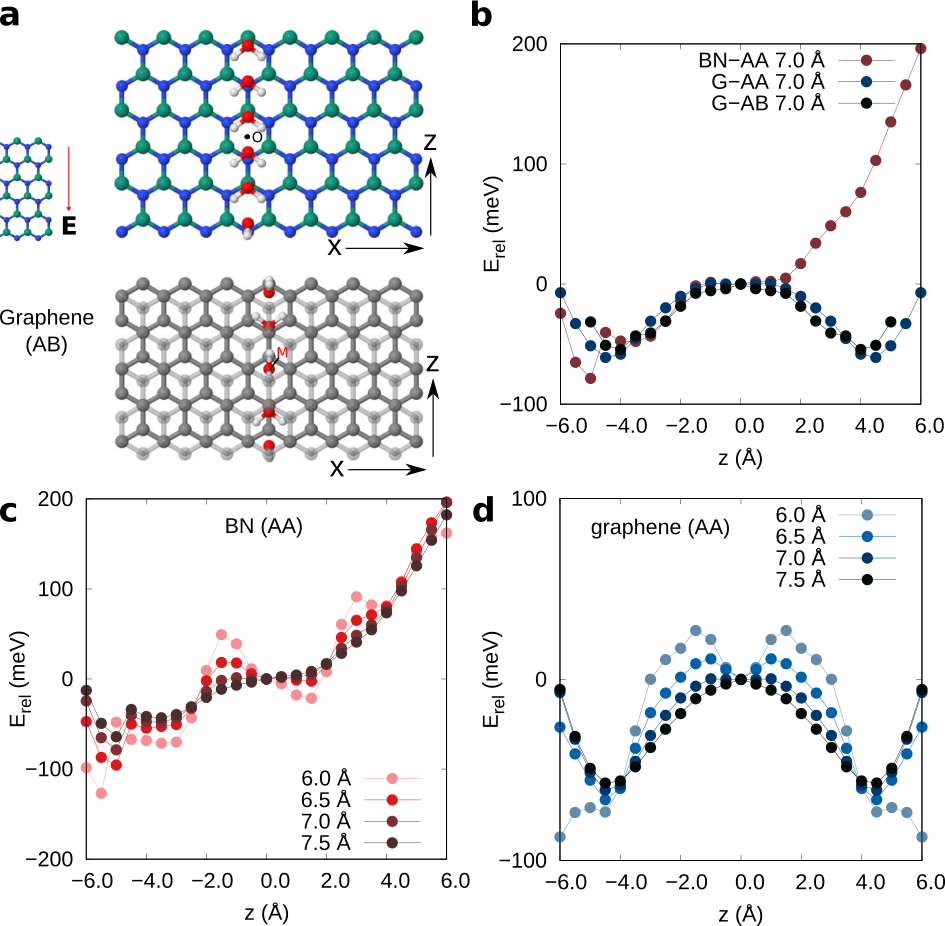}
\caption{Edge effects on the orientation and energy of a single water molecule placed in between $h$-BN or graphene nanoribbons:
(a) Lowest energy configurations of a  water molecule at different $z$, for $D$=7~\AA\ (diagrammatic montage);
only the bottom $h$-BN layer is shown for clarity.
Nitrogen, boron, carbon, oxygen, and hydrogen atoms are represented by blue, green, gray, red and white spheres, respectively.
Periodic boundary conditions are imposed along the $x$ direction {\color{black} (the channel's width)}. ${\bf E}$ indicates the direction of the intrinsic electric field.
(b) Relative energy of a water molecule confined by $h$-BN or graphene in either AA or AB stacking, for an inter-layer spacing of 7~\AA; $z$ is measured from point `O'/`M'.
(c) Relative energy of a water molecule for different inter-layer distances, for AA-stacked $h$-BN;
(d) Relative energy of a water molecule for different inter-layer distances, for AA-stacked graphene.
 \label{fig:mol}}
\end{figure}

For the purpose of illustrating the effect of the $h$-BN polarization on the water, we will use stoichiometric, semi-infinite nanoribbons with asymmetric zigzag edges, i.e. one edge is N-terminated, whereas the opposite one is B-terminated, such as the one represented in Figure~\ref{fig:mol}A. The top and down $h$-BN flakes are stacked in AA configuration. In reality, the channels can be mechanically stacked in many other configurations, but as long as the finite layers are asymmetric with respect to inversion, there will be a finite polarization. 
We start by investigating the potential energy landscape for water molecules confined in such channels.
For comparison, we carried out  similar calculations for graphene, stacked in either AB or AA configurations, both of which have no net polarization.

The nanoribbons used were semi-infinite and, unless otherwise specified, consisted of a total of 168 atoms with a width of 11.3-11.5~\AA~ for the AA-stacked structures, and 140 atoms with a width of 10.0~\AA~for AB-stacked structures. The supercell size used was 17.5$\times$35.0$\times$30.0~\AA$^3$. The relative energies were found to change less than 1~meV when the vacuum spacing was increased by 5~\AA~ along $y$ or $z$.  Spin polarization of the edges was taken into account (for more details, please see the Methods section). 

\subsection{Edge Effects on a Single Water Molecule}
A single water molecule confined between $h$-BN nanoribbons spontaneously reorients itself so that its dipole is  anti-parallel to the polarization of the ribbons. The difference in energy between the parallel and anti-parallel dipoles, for a water molecule placed at equal distance from the two borders (site `O' in Figure~\ref{fig:mol}A) is 0.13 eV and 0.11 eV  for inter-layer distances of 6~\AA\ and 7~\AA, respectively. 
In contrast, in graphene the two orientations are symmetrically equivalent.

However, close to the border of the $h$-BN confining layers, the orientation of the water molecule is no longer dictated by the polarization but rather by the local atomic environment. The energy and orientation of the water molecule at different distances from the border, obtained by fixing the oxygen position to different coordinates along the $z$ axis, are shown in Figure~\ref{fig:mol}A. There are three  effects to have into consideration: (i) the position of the  atoms of the water molecule with respect to the atoms of the lattice of the 2D materials on both sides, (ii) the position of the water molecule with respect to the border and (iii) the orientation of the water dipole with respect to the polarization of the nanoribbon.

A comparison between different stackings and between polar/non-polar NRs shows that the polarization (iii) is the most important factor shaping the potential energy landscape (Figure~\ref{fig:mol}). We define the relative energy of the water molecule, referenced to its energy at $z=0$, as $E_{\rm rel}(z)=E_{\rm tot}(z)-E_{\rm tot}(0) $, where $E_{\rm tot}$ is the total energy calculated using DFT. The respective energy profile is nearly identical for the water molecules confined between AA- or AB-stacked graphene sheets (Figure~\ref{fig:mol}B).

The lowest energy position for adsorption of a water molecule on a single graphene sheet\cite{leenaerts2008adsorption} or in between AA-stacked sheets is at the hexagonal site ('O').  The difference in energy between the hexagonal sites at the centre and at the border is about 70 meV (independently of the inter-layer spacing).

In the case of AA-stacked $h$-BN, the lowest energy position for the oxygen is also close to the centre of the hexagonal ring. For $z<0$, the energy profile resembles that of the water molecule between graphene sheets. However, there is now a strong asymmetry in $z$ due to the polar nature of the $h$-BN nanoribbons. The variations of the potential energy of the molecule across the surface are more pronounced for the smaller inter-layer distances (Figure~\ref{fig:mol}C,D).

The water molecule prefers different orientations with respect to the confining layers, depending on the distance to the border and to the confining layer atoms. 
In the case of confinement by $h$-BN, the water molecule always prefers to be parallel to the confining layers, except when it is close to a border nitrogen atom, in which case it can orient itself perpendicularly to the layers, in order to form hydrogen bonds with the nitrogen atoms. In the case of graphene, however, the water molecule adopts different orientations depending on the closest graphene lattice sites. \textcolor{black}{This difference, as we will see, contributes to the increased order in water phases formed at the 1D$\leftrightarrow$2D transition in $h$-BN channels.}

\begin{figure}[htp]
\centering
\includegraphics[width=8cm]{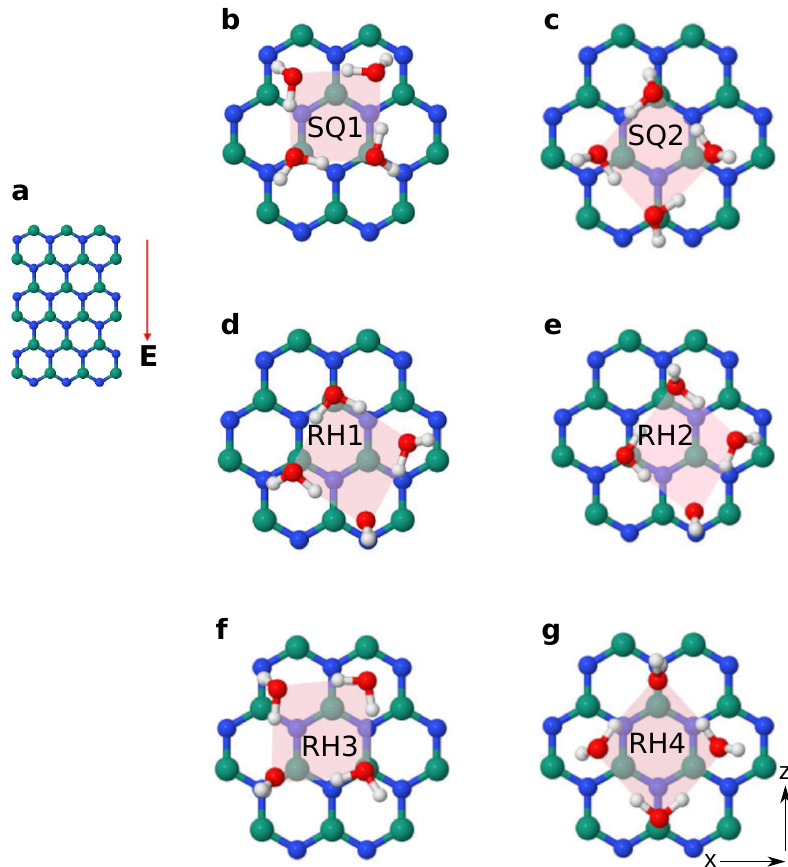}
\caption{Water clusters confined by finite-size $h$-BN nanoribbons.  (a) Orientation of the $h$-BN nanoribbon. The nanoribbons are infinite along the horizontal direction. ${\bf E}$ indicates the direction of the intrinsic electric field. (b-g) Tetra-water clusters in different configurations, see Table \ref{tab:Er} for the relative energies.
The top $h$-BN layer is hidden, for clarity.}
\label{fig:4water}
\end{figure}
\begin{table}[htb]
\caption{Relative energies of tetra-water clusters ($E_{\rm rel}$) confined by $h$-BN or graphene layers, along with the average hydrogen bond energy ($E_{\rm H}$), for an inter-layer distance of 7~\AA. All energies are given in meV.\label{tab:Er}}
\begin{center}
\begin{tabular*}{9.5cm}{lcccc}
\hline
\hline
          &  \multicolumn{2}{c}{$h$-BN (AA)} & \multicolumn{2}{c}{Graphene (AB)} \\
Structure &  $E_{\rm rel}$ & $E_{\rm H}$ &$E_{\rm rel}$ & $E_{\rm H}$\\
\hline
SQ1       & 0    &245 & 1  & 249\\
SQ2       & 5    &253 &  0 & 256\\
RH1         & 8    &149 & 94 & 147\\
RH2       & relaxes to RH1    &149 & 242& 154\\
RH3        & 115  &151 & 244& 154\\
RH4       & 334  &150 &  94& 145\\
\hline\hline
\end{tabular*}
\end{center}
\end{table}

\subsection{Confined Water Clusters}
We now use tetra-water clusters to investigate the relative strength of the water-water interactions. These small water clusters share the same square or rhombic local motifs as the most relevant 2D water phases, a consequence of the maximization of the number of hydrogen bonds between water molecules. \textcolor{black}{However, different from an infinite 2D water layer, water molecules are not all fully bonded, with four hydrogen bonds per molecule.}

Tetra-water clusters are confined by semi-infinite nanoribbons with zigzag edges similar to those described in the previous section. The most relevant configurations are shown in Figure~\ref{fig:4water}.
The clusters SQ1 and SQ2 are nearly planar. In contrast, in the RH1-RH4 clusters, one of the molecules is oriented vertically, forming hydrogen bonds with N, since it is unable to form hydrogen bonds with the other water molecules. Note that this is not common in infinite water layers, as the hydrogen bonding with O is preferred to hydrogen bonding with N, which is less electronegative. 
All the configurations shown have four hydrogen bonds between water molecules, but SQ1 and SQ2, where each water molecule has one and only one H atom involved in a hydrogen bond (Table~\ref{tab:Er}).

\textcolor{black}{
We estimate the average hydrogen bond energy as $E_H=[E_{4{\rm H_2O}}-4E_{1{\rm H_2O}}]/4$, where $E_{4{\rm H_2O}}$ is the energy of the tetra-water cluster, at fixed atomic positions, without the presence of BN, and $E_{1{\rm H_2O}}$ is the energy of a single water molecule. $E_H$ is larger for SQ1 and SQ2, indicating that the larger the number of hydrogen bonds established by a water molecule, the weaker they become. }

Comparing our estimates of the energy scales of the interactions involved in the water layer formation, we find that hydrogen bonds are the strongest, with energies in the range $~$50-250~meV, followed by the dipole reorientation energy ($<$ 110~meV) and border proximity effects ($~$100 meV). Comparatively, the local changes of potential energy  due to proximity to the B,N/C atoms far from the border, is small ($<$50~meV), specially for inter-layer distances of 7.0-7.5~\AA.
Taking this into account, it is not surprising that 2D water forms water phases that maximize the formation of hydrogen bonds. 
At room temperature, it favours square and rhombic phases, instead of forming phases commensurate with the confining lattice.
\textcolor{black}{In the following sections, we will study such 2D water phases and how they change at the 2D$\leftrightarrow$1D transition.}

\section{Phases of Confined 2D Water\label{MD}}

We used classical molecular dynamics simulations to explore the phase diagram of 2D water at 300~K.
AA-stacked $h$-BN layers and AB-stacked graphene layers were chosen as model systems. 
The phase diagrams are calculated at room temperature,
for different distances between plates ($D$) and lateral pressures.
The molecular dynamics simulations are performed at constant number of particles ($N$), with water covering given area $A$, determined by the size of the $h$-BN supercell.  For a given $N$, we calculate the pressure on the (virtual) supercell walls perpendicular to 
the walls separation, $P(D)$. Phase transitions between the different phases accessible to the system can be identified by a non-monotonic change of $P(D)$ with $D$ at constant $N$~\cite{gao2018phase}.
{\color{black}The procedure is similar to what has been used in  \cite{gao2018phase}, and our outcomes are consistent with previous results for water confined by graphene 
layers (see Supporting Information~S1).}
We focus mostly on the regions of stability of monolayer water, in either the solid or liquid state, and under compression.
Further details about the molecular dynamics simulations can be found in the Methods section. 

A comparison between the classical potentials and DFT for the system studied in the previous section can be found in \textcolor{black}{ Supporting Information~S2. While the classical potentials perform poorly when comparing the energies of water clusters with on or two hydrogen water molecules, they perform better for fully bonded water phases. We have chosen the classical potential SPCE\cite{berendsen1987missing}.
This model gives the best structural parameters and self-diffusion constant among the most commonly used nonpolarized classical water models\cite{mark2001structure}. The impact of the classical water model choosen will be discussed in more detail in the next sections.}

\begin{figure}[htp]

\centering
\includegraphics{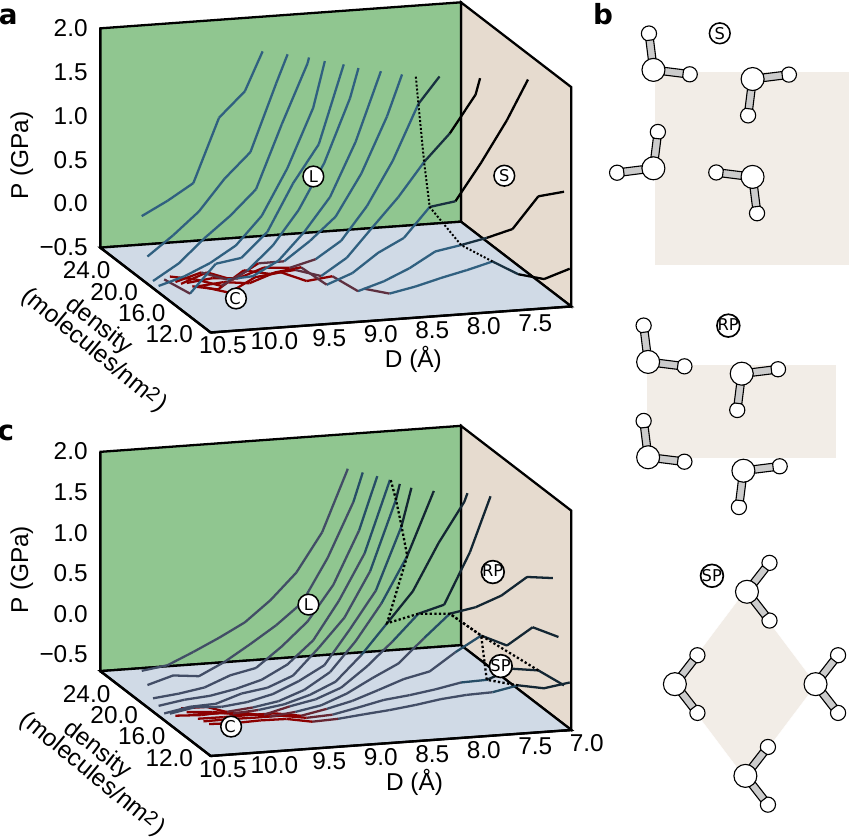}
\caption{Phase diagrams for water confined by either (a) infinite $h$-BN sheets or (c) finite $h$-BN nanoribbons.
The liquid, non-polar square, rhombic polar and square polar phases are designated by `L', `S', `RP', and `SP', respectively. `C' indicates cavitation. 
(b) Schematic illustration of the crystalline water phases (not to scale).}
\label{fig:phasediag}

\end{figure}

Infinite boron-nitride sheets, modelled using periodic boundary conditions along both in-plane directions, have no net intrinsic electric field, and the phases of water confined by such sheets are identical to those found when water is confined by graphene (Figure~\ref{fig:phasediag}A). The main crystalline phase encountered is the square phase (S), shown in Figure~\ref{fig:phasediag}B, which has been  observed in previous experiments\cite{algara2015square} and predicted theoretically\cite{zangi2003monolayer,gao2018phase,zhu2015compression,jiang2021first,neek2016commensurability,qiu2015water}.
However, the situation is different when water is confined by polar BN nanoribbons.

\subsection{Between 2D and 1D: Confinement by $h$-BN Nanoribbons\label{MD-nr}}

We now consider 2D water nanoribbons, which can be viewed as a quasi-1D system, obtained by confining the 2D water in between $h$-BN nanoribbons.
 These ribbons, in the $xOz$ plane, are finite along $z$ and infinite along $x$. {\color{black} The channel's length ($L$) is approximately 42~\AA. 
The supercell is 65.1 and 63.9~\AA\ wide (the channel's width $W$) along the $x$ direction for $h$-BN and graphene, respectively.}
The water is kept in the confined region by perfectly reflective wall boundaries placed parallel to the nanoribbon edges.

\begin{figure*}[htp]
\centering
\includegraphics[width=\textwidth]{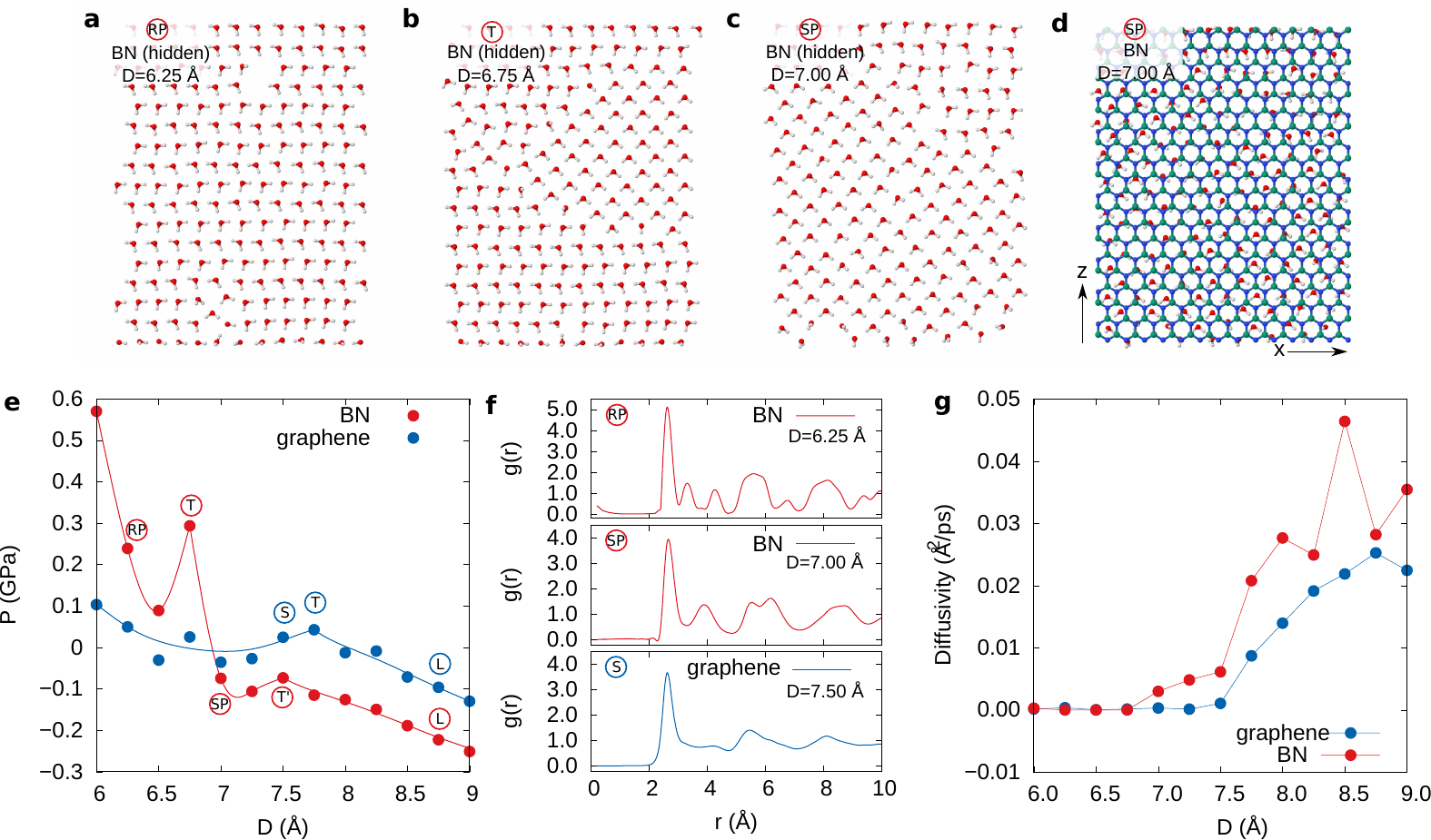}
\caption{
\label{fig:sigma12}
Phases of 2D water confined by semi-infinite monolayer $h$-BN or graphene nanoribbons. Structures for $h$-BN confinement for (a) rhombic polar phase, (b) rhombic polar and square polar phases coexisting at transition point, (c) square polar phase, and (d) square polar phase, showing the orientation of the ribbon. The respective interlayer distances $D$ are shown in the insets;
(e) phase diagram; (f) 2D radial distribution functions ($g(r)$); and (g) 2D diffusivity obtained from the mean square displacements.  The density is 12.4 molecules/nm$^2$ for graphene and 12.2 molecules/nm$^2$ for $h$-BN.}
\label{fig:ribbonBN}
\end{figure*}

Two-dimensional water confined between $h$-BN 
nanoribbons shows two distinct solid phases. These are a rhombic polar phase (RP) and a square polar (SP) phase. For a density of about 12 molecules/nm$^2$ (the density of the 2D square ice observed experimentally\cite{algara2015square}), the transition between the two solid phases occurs at $D=6.75$~\AA.
Below $D=6.50$~\AA, there is a  rhombic phase where the oxygen atoms arrangement shows rhombus motifs  (Figure~\ref{fig:ribbonBN}A).
This is the same as the rhombic structure proposed in\cite{chen2016two}, except that we did not observe the duplication of unit cell at room temperature. The electrical dipoles of the O-H bonds are approximately parallel  to the $z$ direction pointing to the N-terminated edge, such that the net polarization of the water layer is opposed to that of the $h$-BN nanoribbon. 
At $D=6.75$~\AA, there is a phase transition, with domains of the two phases coexisting (Figure~\ref{fig:ribbonBN}B).
Above this transition point, the square polar phase, where  the electric dipole of each water molecule is aligned with the $-z$ direction, becomes the most stable (Figure~\ref{fig:ribbonBN}C,D). This phase is the same as the `PR' (puckered rhombic) phase proposed in\cite{li2019two}. Besides, in all regions of the phase space, there are structure changes near the N-terminated zigzag edge.

Water confined by the graphene nanoribbon, in contrast, only shows a solid-liquid transition at $D=$7.75~\AA (Figure~\ref{fig:ribbonBN}E). 
This is similar to what has been previously found for 2D water confined by infinite graphene planes\cite{gao2018phase} (see Supporting Information~S1).

The phase transitions give rise to anomalies in the $P(D)$ plots (Figure~\ref{fig:ribbonBN}E). Additionally, different phases have distinct radial distribution functions (RDFs) (Figure~\ref{fig:ribbonBN}F).
In the rhombic polar structure, each oxygen atom has four nearest neighbours at 2.6-2.8~\AA, and two second nearest neighbours at 3.2-3.4~\AA. In contrast, in the square polar structure, each oxygen atom also has four nearest neighbours at 2.6-2.8~\AA, but there is a second distinct peak due to the four second nearest neighbours at 3.8-3.9~\AA. In both the rhombic and square polar phases of water confined by the $h$-BN ribbons, the peaks of the RDF are sharper than those of the RDF for water confined by graphene. 
A phase diagram for varying molecular density and confining inter-layer distance is shown in Figure~\ref{fig:phasediag}B.

The diffusivity is also different for the rhombic and square polar phases, being  higher for  the latter, and it dramatically increases in the liquid state (Figure~\ref{fig:ribbonBN}G). Thus, despite some degree of disorder, especially in the H sublattice, it is clear that all the three phases observed (square, rhombic polar and square polar) are solid at room temperature.

\begin{figure}[htp]     
\centering
\includegraphics[width=0.5\textwidth]{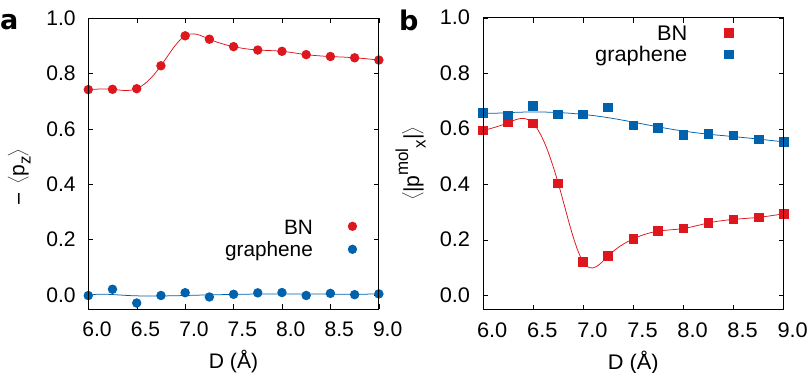}\\
\caption{Averaged dipole moment per water molecule, in units of the water dipole (2.35 D for SPCE), for 2D water confined by $h$-BN and graphene nanoribbons:
(a) average polarization of the $z$-component and (b) average of the absolute value of the $x$-component of the molecular polarizations.}
\label{fig:polribbons}
\end{figure}

In order to quantify the presence of the polar phases for the 2D water, we have calculated the average electric dipole per water molecule, $\langle {\bf p}\rangle$ (Figure~\ref{fig:polribbons}).
This is zero for the square water phase, but non-zero for the rhombic polar and square polar phases.

For the water confined by the $h$-BN nanoribbons, there is a noticeable increase of the value of $\langle {p_z}\rangle$ towards its maximum value at the phase transition from the rhombic polar phase to the square polar phase (Figure~\ref{fig:polribbons}A). Even the liquid phase is highly polarized. In contrast, $\langle p_z\rangle$ is zero for the water confined by the graphene nanoribbon.

Further, to help differentiating between the rhombic polar and the square polar phases, we calculated $\langle|p_x^{\rm mol}|\rangle$, the ensemble average of the absolute values of the molecular polarization component along $x$. This is zero for the perfect square polar structure, since all the molecules have a mirror symmetry plane parallel to $xOy$, but it is non-zero for the rhombic polar structure. Thus, an abrupt decrease of $\langle|p_x^{\rm mol}|\rangle$ is observed at the phase transition from the rhombic polar phase to the square polar phase (Figure~\ref{fig:polribbons}B). In contrast, the value of $\langle|p_x^{\rm mol}|\rangle$ remains constant for graphene.

\textcolor{black}{
We note that the stabilisation of polar 2D water phases in the case of $h$-BN confinement is not an artifact of the choice of classical potential since, according to our tests for strictly planar 2D water phases, in Supporting Information~S2.2, this model slightly penalizes the polar water phases, compared to DFT. In Supporting Information S3.1, we show that the presence of polar solid phases and the solid to liquid transition near $D\sim 7$\AA\ are independent of the choice of classical potentials.}

\subsection{Electric Field Variation within the $h$-BN Nanoribbon Width  \label{E}}

We have shown that water confined by finite $h$-BN nanoribbons acquires new phase states which are not present in water confined by infinite $h$-BN layers.
We have attributed this difference to the macroscopic polarization of the nanoribbon system associated with an intrinsic electric field. The energy of a water molecule in the resulting electric field, $E_{\rm dip}$, is given by $E_{\rm dip}=-{\bf p}\cdot{\bf E}=-pE_z\cos\theta$, where {\bf p} is the electric dipole of the water molecule, {\bf E} is the electric field,  which is assumed to be along the $z$ direction, ${\bf E}=E_z{\bf e}_z$, and $\theta$ is the angle between {\bf p} and {\bf E}. We now analyse quantitatively the electric field produced by the nanoribbon, and how it varies in-plane and within the nanoribbon width.

\begin{figure}[htp]
\centering
\includegraphics[width=8cm]{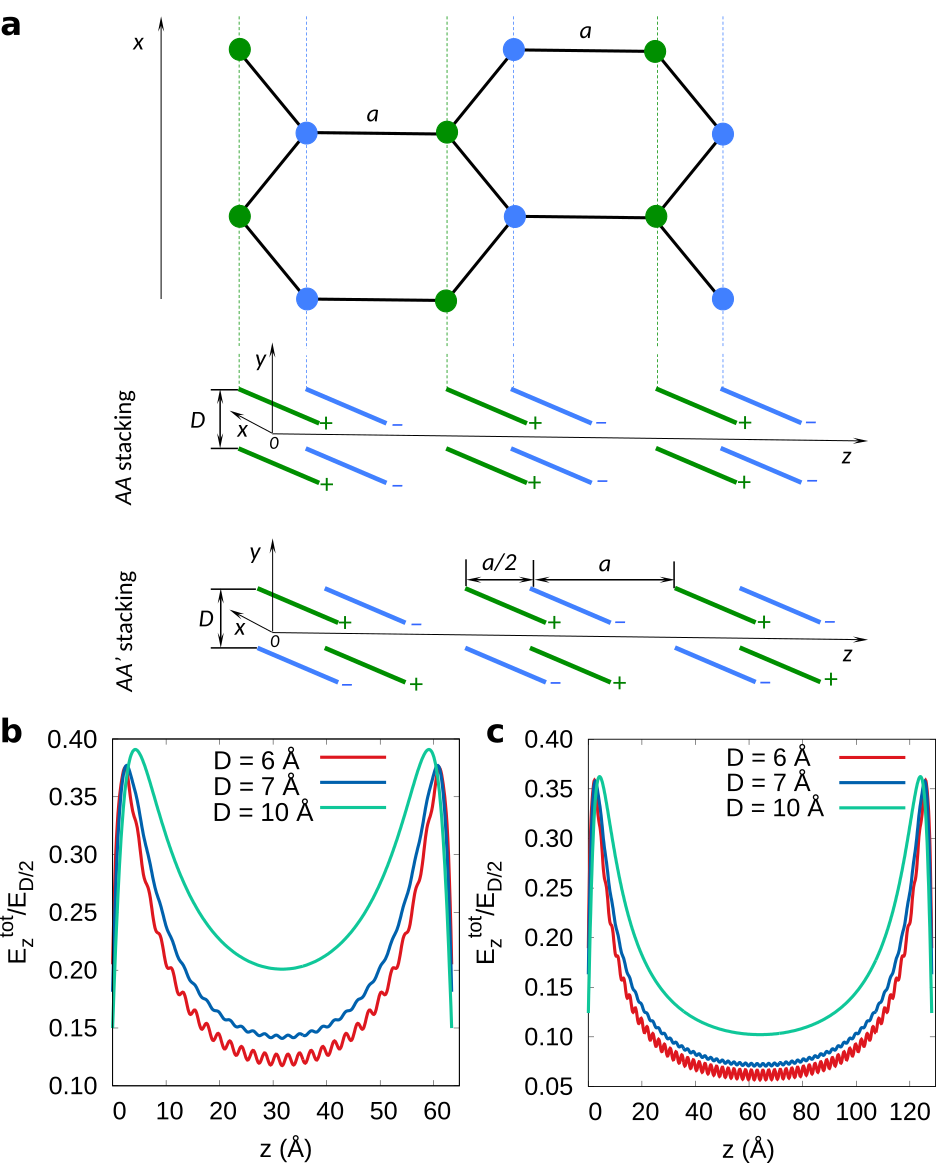}
\caption{Effective electrostatic model for the in-plane mid-gap electric field produced by the polar $h$-BN nanoribbons.
(a) Difference between AA and AA$^\prime$ stacking assemblies. The boron and nitrogen linear charge distributions are shown in green and blue, respectively. 
(b,c) In-plane mid-gap electric field given by Eq. (\ref{tot-field}) in units of $E_{D/2}=\lambda/(\pi \epsilon D)$ 
for AA stacking with (b) $L\sim$65~\AA (30 unit cells) and (c) $L\sim$130~\AA (60 unit cells).
The in-plane field at the middle of the gap vanishes in the case of AA$^\prime$ stacking.}
\label{fig:texture}
\end{figure}

It is possible to quantify the electric field in $h$-BN channels
by means of an equivalent electrostatic model, Figure~\ref{fig:texture}A, representing the boron and nitrogen edges as linear charge distributions,
with linear charge density $\lambda$ of the order of 0.1 $e$/\AA.
The electric field of a single line placed at $z=z_0$, $y=\pm D/2$ reads
\begin{equation}
{\mathbf E}=\frac{\lambda}{2\pi \epsilon} \frac{(z-z_0)\mathbf{e}_z + (y\mp D/2)\mathbf{e}_y}{(z-z_0)^2 + (y\mp D/2)^2}.
\label{line-field}
\end{equation}
where $\mathbf{e}_{y,z}$ are the unit vectors, and $\epsilon$ is the dielectric permittivity.
Since the water flow is restricted to the plane, the water molecules are most sensitive to the in-plane electric field (along $\mathbf{e}_z$) in the middle of the interlayer gap ($y=0$).
The total in-plane electric field at mid-gap in the case of AA stacking can be written as
\begin{eqnarray}
\nonumber E_z^\mathrm{tot}&=&\frac{\lambda}{\pi \epsilon} \sum\limits_{n=0}^{N-1}\left[\frac{z-n l}{(z-nl)^2 + (D/2)^2} \right. \\
&& \left. - \frac{z-n l-z_0}{(z-nl-z_0)^2 + (D/2)^2}\right],
\label{tot-field}
\end{eqnarray}
where $l=3a/2$, $z_0=a/2$ with bondlength $a=1.44$ \AA \cite{yamanaka2016energetics}, and $N$ is the number of $h$-BN unit cells along the channel.
The $n=0$ term corresponds to the field produced by two boron edges with $+\lambda$ placed at $z=0$, $y=\pm D/2$ and 
two nitrogen nearest-neighbour lines with $-\lambda$ placed at $z=z_0$, $y=\pm D/2$.
The $n=N-1$ term corresponds to the field produced by two nitrogen edges placed at $z=3aN/2-a$, $y=\pm D/2$
and two boron nearest-neighbour lines placed at $z=3a(N-1)/2$, $y=\pm D/2$.
Both the upper and lower $h$-BN layers are thus boron-terminated at $z=0$ and nitrogen-terminated at $z=3aN/2-a$.
The system as a whole is electrically neutral.
The field contributions produced by the upper and lower $h$-BN layers 
are equivalent in the middle of the interlayer gap and sum-up to a finite value.

The total in-plane mid-gap electric field is maximal near the edges and decreases in the middle of the channel (Figure~\ref{fig:texture}B,C).
However, it remains comparable with the edge field if the ribbon width is of the order of 10 \AA.
The absolute value of the field at the channel edges is about $\lambda/(\pi \epsilon D)\sim 10^9$ V/m.
It is strong enough to fully polarize water molecules \cite{booth1951dielectric}.
The field oscillates as a function of the $z$ coordinate, with amplitude depending on the ratio between $a$ and $D$.
If $a/D\ll 1$, then oscillations are negligible.

\begin{figure}[!thb]
\centering
\includegraphics{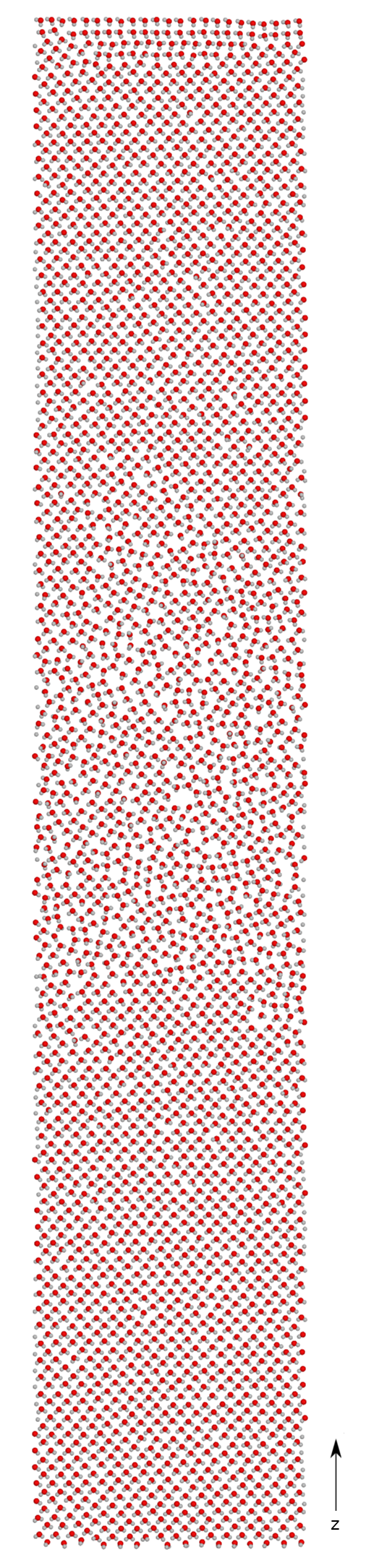}\\
\caption{ Snapshot of the water monolayer in a \textcolor{black}{$h$-BN channel with width $L=368$~\AA\, along the $z$ direction, and $D=7$~\AA, along the direction perpendicular to the plane.
The channel is infinite along the horizontal direction. } The water density is 12.4 molecules/nm$^2$.
Transition between ordered and disordered states occurs at the distance of about 100~\AA~ from the nearest edge
due to the vanishing in-plane mid-gap electric field far away from the nanoribbon's edges, see Figure~\ref{fig:texture}.
\label{fig:long}
}
\end{figure}

In an MD simulation for water confined by $h$-BN nanoribbons of a longer length ($L$=368~\AA), we are able to see the ordered, polar phase within 100~\AA~of the edges, while at the centre of the water layer the electric field vanishes, and the water layer becomes disordered (Figure~\ref{fig:long}). Near the centre of the water layer, the 2D water is still polarized, although less than near the edges (see Supporting Information S4).

In the case of  AA$^\prime$ stacking (with top and bottom edges aligned) the situation is completely different because each term originating from a line at $y=+D/2$ is
explicitly compensated by the term with $y=-D/2$ as having an opposite sign.
Hence, the total in-plane mid-gap electric field vanishes in the case of  AA$^\prime$ stacking.

From this model, we can also see that if $L\to\infty$, then the average electric field tends to zero, 
in agreement with the molecular dynamics results shown earlier for the infinite $h$-BN sheet. This is therefore a nanoscale effect.

\subsection{Dynamics of Confined Water Flow in Polar BN Channels\label{MD-flow}}

\begin{figure}[htp]
\centering
\includegraphics{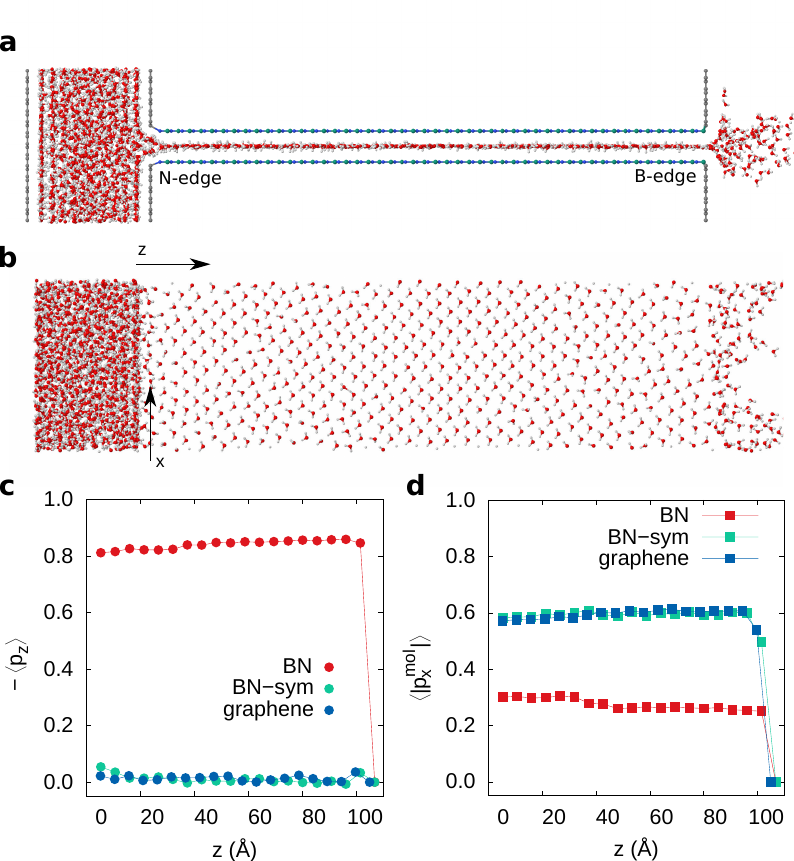}\\
\caption{Structure of confined water flowing between two reservoirs with a pressure difference of $\Delta\mathcal{P}=10$~atm: 
(a) system diagram, (b) snapshot of the water structure, (c) average of the $z$-component of dipole moment per water molecule,
in units of the water dipole (2.35 D for SPCE), and (d) average of the absolute value of the $x$-component of the molecular polarizations, for graphene, $h$-BN (AA), and for a centrosymmetric channel with AA$^\prime$ stacking (BN-sym). The channel inter-layer distance is 6~\AA.}
\label{fig:flow}
\end{figure}

Finally, we have considered the phases of water during stationary flow in the presence of a pressure gradient.
The model system is represented in Figure~\ref{fig:flow}A. It consists of two reservoirs separated by a 2D channel with length $L$=105~\AA~ for graphene and $L$=108~\AA~ for $h$-BN, corresponding to the same number of repeating units. The reservoir on the left has a moving graphene piston with a total applied force $f=PA$, where $A$ is its area.
The channel consisted of either (i) two graphene sheets or (ii) two $h$-BN sheets with AA stacking and N-terminated edge on the $+z$ side, B-terminated edge on the $-z$ side. The inter-layer distances were 6~\AA~ and 7~\AA. 

We considered pressure differences between the two reservoirs from 5 to 208 atm.
In all cases, the water layer was in the solid state with a structured oxygen sublattice,
but the O-H bonds were polarized for the water flow through the $h$-BN channel.
This is illustrated in Figure~\ref{fig:flow}C,D for a channel interlayer distance of 6~\AA\  and  a pressure difference of 10.4 atm. The density was in the range 10.8-14.9 molecules/nm$^2$ for graphene and 11.7-12.4 molecules/nm$^2$ for $h$-BN. The value of  $\langle {p_z}\rangle$ is close to the maximum polarization for $h$-BN, but nearly zero for graphene. There is a mixture of rhombic polarized and square polarized phases (the latter featured in Figure~\ref{fig:flow}A). This is confirmed by the $\langle|p_x^{\rm mol}|\rangle$ value, which lies between the values characteristic of these two phases.
This is in contrast with the graphene confinement, for which there is no net water polarization.

\section{Conclusions}
Density functional theory calculations, together with molecular dynamics simulations, indicate that 2D water confined by finite $h$-BN nanoribbons is more ordered than confined by identical graphene nanoribbons. The difference is mainly due to the presence of an intrinsic electric field in finite, asymmetric $h$-BN layers, and is not observed for water confined by infinite $h$-BN layers which have no net electric field. The reorientation energy of the molecules is higher close to the edges than at the centre of the nanoribbon, 
and its average value over the whole surface vanishes as the nanoribbon width tends to infinity.

At room temperature, water confined by AA-stacked, polar $h$-BN layers less than 7.5~\AA~ apart is crystalline and can assume one of two polar phases\textcolor{black}{, RP and SP. The regions of confining distance where RP or SP dominate depend on the classical potential employed}. When flowing due to a pressure gradient a strong polarization of the water molecules is maintained during flow.

The qualitative results obtained here can be generalized for nanoribbons of different nanoscale widths, as long as the polarity of the top and bottom layers is the same. If they have inversion symmetry, however, no difference is expected as in the case of graphene.

The edges of the nanoribbons give us the ability to control the crystalline states of water confined in between. This approach can be further explored 
considering different polar materials. Even though an external electric field can change the water structure and dynamics in nanotubes \cite{ritos2016electric,yamamoto2017water,winarto2015structures,su2011control} and inorganic systems\cite{bratko2009water}, 
the method could be hazardous in practice. In contrast, making use of the intrinsic electric field in $h$-BN, which is already a typical material used in prototypes of nanodevices and  nanocapillaries, opens the way to safe electrical manipulation of the water structure.
\textcolor{black}{The main difficulty at present is the production of BN with clean zigzag edges. Such edges have been sparsely reported even for graphene\cite{wu-PhysRevLett.120.216601}. The process of edge creation would have to be compatible with the methods currently used for the
creation of nanochannels by assembly of laminar 2D sheets. Nevertheless, the polar 2D water phases described here can potentially be found in regions of water intentionally or unintentionally intercalated between BN sheets during the assembly of heterostructures.
}

\section{Methods}
\subsection{DFT Calculations}
First-principles DFT calculations were carried out using the {\sc Quantum ESPRESSO} package\cite{giannozzi2017advanced}. 
The VDW-DF1 exchange and correlation energy functional was used\cite{dion2004van}.
This exchange functional has been previously tested for hydrogen-bonded systems with accurate results.\cite{Langreth_2009} Ultra-soft pseudo-potentials were used to account for the core electrons\cite{rkkjus}.
We employed a plane wave basis set with kinetic energy cutoffs of 35~Ry to describe the electronic wave functions. 
The Brillouin zone was sampled using a $\Gamma$-centered 4$\times$4$\times$1 Monkhorst-Pack (MP) grid\cite{mpgrid} for all calculations.

\subsection{Classical Molecular Dynamics}
The classical molecular dynamics simulations were performed using the {\sc lammps} code \cite{plimpton1995fast}.

\textcolor{black}{
Many classical potentials for water have been proposed in the literature. We have employed the reparameterized simple point charge model (SPCE) model \cite{berendsen1987missing}, for which water-BN interactions have been determined by Wagemann et al. (SPCE/WWDM)\cite{wagemann2020wetting}. Alternatives are TIP4P and TIP3P, for which water-BN interactions have been determined by Wu  et al. (WWA)\cite{wu2016hexagonal} and Hilder et al. (HYGGBRC)\cite{hilder2010validity}, respectively. We could not test TIP5P because we did not find BN interaction parameters for this potential in the literature.}
In all cases the water-carbon interaction was modeled by a Lennard-Jones potential between oxygen and carbon atoms.

\textcolor{black}{
We employed the SPCE water model, which is known to give very good structural parameters and self-diffusion constant for 3D water \cite{mark2001structure}. The SPCE water model has also previously been used to study the phase diagram of 2D water using the method that we adopted in the present paper, with results consistent with experiment.\cite{gao2018phase} We found this model to be in reasonable agreement with DFT for strictly planar 2D water (Supporting Information Section S2.2). 
All the three potentials considered favour SQ with respect to RP for strictly 2D water, however by a small energy margin (Supporting Information S2).
Thus, the disappearance of the non-polar SQ phase from the phase diagram of water confined by BN nanoribbons cannot be attributed to limitations of the classical potential. A comparison of the main results of this paper for different classical potentials can be found in Supporting Information S3.
}

\textcolor{black}{The water-BN interaction parameter of Wageman et al.\cite{wagemann2020wetting} employed here consists of a Lennard-Jones potential between oxygen and boron or nitrogen. These result in a contact angle of 73$^\circ$. The results for different water-BN interacting potentials were found to be consistent (see Supporting Information S3), except when using the potential of Hilder et al.\cite{hilder2010validity}, due to the large N and B charges in the latter ($\pm$0.98), which are larger than previous estimates ($<0.3$) for similar structures\cite{martinez2014atomistic}.}

Long-range Coulomb forces were computed using the particle-particle particle-mesh (PPPM) method.
Water molecules were kept at a constant temperature of 300 K using a Nos\'e-Hoover thermostat with a damping constant of 10~fs (100 timesteps). The graphene or boron nitride atoms were not coupled to the thermostat.
A timestep of 0.1 fs was used.

\begin{acknowledgement}
This research / project is supported by the Ministry of Education, Singapore, under its Research Centre of Excellence award to the Institute for Functional Intelligent Materials, National University of Singapore. The computational work was supported by the Centre of Advanced 2D Materials (CA2DM), funded by the 
National Research Foundation, Prime Ministers Office, Singapore, under its Medium-Sized Centre Programme.
M.T. thanks the Director’s Senior Research Fellowship from CA2DM [R-723-000-001-281] for support.
\end{acknowledgement}

\begin{suppinfo}

The following files are available free of charge.
\begin{itemize}
  \item MD Simulation of 2D Water Confined by Infinite Boron Nitride Sheets
\end{itemize}

\begin{itemize}
  \item Comparison of DFT and Classical Potentials: Water Clusters
\end{itemize}

\begin{itemize}
  \item\textcolor{black}{Comparison between Classical Potentials: Phase Diagram for Water Confined by BN NRs}
\end{itemize}

\begin{itemize}
  \item Structure of 2D Water Confined by Graphene NRs
\end{itemize}

\begin{itemize}
  \item\textcolor{black}{Wide} BN Nanoribbon Simulation
\end{itemize}

\end{suppinfo}

\bibliography{main.bib}

\end{document}